\documentclass[aps, prd, twocolumn, floatfix, showpacs, nofootinbib,
               longbibliography]{revtex4-1}

\usepackage{amsmath}
\usepackage{amssymb}

\usepackage[]{graphicx}
\usepackage{subfig}

\usepackage{color}

\usepackage{natbib}

\usepackage{placeins}

\usepackage{multirow}

\usepackage{hhline}

\usepackage{acronym}

\usepackage{hyperref}
\hypersetup{
    colorlinks,
    allcolors={blue!60!black}
}

\usepackage{esint}

\newcommand{\fdot}{\dot{f}}
\newcommand{\F}{{\mathcal{F}}}
\newcommand{\Fhat}{\widehat{\F}}
\newcommand{\rhohat}{\widehat{\rho}}
\newcommand{\twoFhat}{\widehat{2\F}}
\newcommand{\Ftilde}{\widetilde{\F}}
\newcommand{\rhotilde}{\widetilde{\rho}}
\newcommand{\twoFtilde}{\widetilde{2\F}}
\newcommand{\A}{\boldsymbol{\mathcal{A}}}
\newcommand{\blambda}{\boldsymbol{\mathbf{\lambda}}}
\newcommand{\Signal}{^\mathrm{s}}

\newcommand{\Tspan}{T}
\newcommand{\Tcoh}{T_{\rm coh}}

\newcommand{\Nseg}{N_\mathrm{\rm seg}}

\newcommand{\Nburn}{N_\mathrm{\rm burn}}
\newcommand{\Nprod}{N_\mathrm{\rm prod}}
\newcommand{\Ntemps}{N_{\temp}}
\newcommand{\Nsft}{N_\mathrm{SFT}}
\newcommand{\Nwalkers}{N_{w}}

\newcommand{\temp}{t}

\renewcommand{\H}{\mathcal{H}}
\newcommand{\Hs}{\H_{\rm S}}
\newcommand{\Hg}{\H_{\rm G}}
\newcommand{\data}{\boldsymbol{x}}
\newcommand{\noise}{\boldsymbol{n}}
\newcommand{\signal}{\boldsymbol{s}}

\newcommand{\Bsg}{B_{\rm S/G}}

\newcommand{\Sn}{S_{\rm n}}

\newcommand{\rhohatmax}{\hat{\rho}_{\mathrm{max}}}

\newcommand{\N}{\mathcal{N}}
\newcommand{\Neff}{\mathcal{N}^*}
\newcommand{\tildegrss}{\widetilde{g}^\mathrm{rss}}
\newcommand{\hatgrss}{\widehat{g}^\mathrm{rss}}
\newcommand{\Depth}{\mathcal{D}}
\newcommand{\noiseUnits}{Hz$^{-1/2}$}

\newcommand{\tauS}{\tau^{(0)}_\mathrm{S}}
\newcommand{\tauSCeff}{\tau^{(0)}_\mathrm{C}}
\newcommand{\tauCore}{\tau^{(0)}_\mathrm{core, LD}}
\newcommand{\tauBuffer}{\tau^{(0)}_\mathrm{buffer, LD}}
\newcommand{\tauLD}{\tau^{}_\mathrm{LD}}
\newcommand{\tauT}{\tau^{(0)}_\mathrm{T}}
\newcommand{\tauFeff}{\tau_{\mathcal{F}}^{\mathrm{eff}}}

\def\AllSkyMCNoiseOnlyMaximum{58.4}
\def\BasicExampleDays{100}
\def\BasicExampleDeltaFone{1.8{\times}10^{-13}}
\def\BasicExampleDeltaFzero{2.0{\times}10^{-7}}
\def\BasicExampleDepth{10.0}
\def\BasicExampleFailNstar{10^{7}}

\def\BasicExampleNstar{10}
\def\BasicExampleSqrtSn{1.0{\times}10^{-23}}
\def\BasicExamplehzero{1.0{\times}10^{-24}}

\def\BasicExamplenwalkers{100}
\def\DirectedFollowUpDepth{40.0}

\def\DirectedFollowUpNstarMax{1000}
\def\DirectedFollowUpSqrtSn{1.0{\times}10^{-23}}
\def\DirectedFollowUphzero{2.5{\times}10^{-25}}

\def\DirectedMCNoiseOnlyMaximum{50.9}

\def\GridedFrequencyDelta{0.50}
\def\GridedFrequencyInitialSqrtSx{$1.0{\times}10^{-23}$}
\def\GridedFrequencyInitialT{100}
\def\GridedFrequencyInitialhzero{$1.7{\times}10^{-25}$}

\def\GridedFrequencyT{5}
\def\GridedFrequencycosi{0.10}

\def\SCDirectedFUDeltaFone{1.8{\times}10^{-11}}
\def\SCDirectedFUDeltaFzero{2.0{\times}10^{-5}}

\newacro{MCMC}{Markov chain Monte Carlo}
\newacro{MC}{Monte Carlo}
\newacro{ACT}{autocorrelation time}
\newacro{CW}{continuous gravitational wave}
\newacro{SNR}{signal-to-noise ratio}

\usepackage[normalem]{ulem}	%% only added for 'strikeout' \sout
\usepackage[usenames,dvipsnames]{xcolor}

\newcommand{\addtext}[1]{\textcolor{black}{#1}}
\newcommand{\meta}[1]{\addtext{#1}}
\newcommand{\CHECK}[1]{\textcolor{black}{#1}}

\begin{document}

\title{Hierarchical multi-stage MCMC follow-up of continuous gravitational wave candidates}

    \author{G. Ashton}
    \email[E-mail: ]{gregory.ashton@ligo.org}
    \affiliation{Max Planck Institut f{\"u}r Gravitationsphysik
                 (Albert Einstein Institut) and Leibniz Universit\"at Hannover,
                 30161 Hannover, Germany}
    \author{R. Prix}
    \affiliation{Max Planck Institut f{\"u}r Gravitationsphysik
                 (Albert Einstein Institut) and Leibniz Universit\"at Hannover,
                 30161 Hannover, Germany}

\date{\today}

\begin{abstract}

Leveraging \ac{MCMC} optimization of the $\mathcal{F}$-statistic, we introduce
a method for the hierarchical follow-up of continuous gravitational wave
candidates identified by wide-parameter space semi-coherent searches. We
demonstrate parameter estimation for continuous wave sources and develop a
framework and tools to understand and control the effective size of the parameter space,
critical to the success of the method. Monte Carlo tests of simulated
signals in noise demonstrate that this method is close to the theoretical
optimal performance.

\end{abstract}

%\pacs{04.80.Nn, 97.60.Jd, 04.30.Db}
%\input{git_tag.tex}
%\date{\commitDATE; \commitIDshort-\commitSTATUS, \dcc P1700455}

\maketitle

\section{Introduction}

A target for the advanced gravitational wave detector network of LIGO
\citep{aasi2015advanced}
and Virgo \citep{acernese2014advanced} are long-lived quasi-periodic \acp{CW}
from neutron stars. Detection of such signals
would provide unique astrophysical insights and has hence
motivated numerous searches \citep{abbott2017knownpulsar, abbott2017scox1markov,
abbott2017scox2cross, abbott2017allsky, abbott2017allskyeinstein, abbott2017narrowband}.

The gravitational wave signal strain from a rotating neutron star $h(t, \A, \blambda)$
\citep{jks1998} has two distinct sets of parameters:
$\A{=}\{h_0, \cos\iota, \psi, \phi_0\}$, a set of the four
\emph{amplitude parameters} and
$\blambda$, a set of the \emph{phase-evolution parameters} consisting of the
sky-location, frequency $f$, any spin-down terms $\{\dot{f}, \ddot{f},
\ldots\}$ and binary orbital parameters, if required (see \citet{prix2009gravitational}
for a review).

CW searches are often based on a fully coherent matched-filtering method
whereby a \emph{template} for $h(t; \A, \blambda)$ is convolved against the
data, resulting in a detection statistic. Four search categories can be identified
dependent on the level of prior knowledge about certain parameters:
\emph{targeted} searches for a signal from a known pulsar where
the phase-evolution parameters are considered known; an extension to targeted
searches are \emph{narrow-band} searches which allow some uncertainty in the
frequency and its derivatives; \emph{directed} searches
in which the sky-location is considered known, but not the other phase evolution
parameters (i.e., searching for the neutron star in a supernova remnant which
does not have a known pulsar); and \emph{all-sky} searches where none of the
phase-evolution parameters are known. Unknown parameters must
be searched over either numerically using a template bank, or
analytically maximized.

Wide-parameter space searches (directed or all-sky) often use a \emph{semi-coherent} search because at fixed
computing cost it is typically more sensitive to unknown signals than a fully coherent
search \citep{brady1998, prix2012, prix2009gravitational}.  Generically speaking, a semi-coherent
search involves dividing the total observation span $\Tspan$ into $\Nseg$
segments, in each segment computing a fully coherent detection statistic, then
recombining the detection statistic in each segment into a semi-coherent
detection statistic; for details of specific variations on this principle, see
\citep{brady2000, krishnan2004, astone2014, allyskyS42008, pletschallen2009,
pletsch2008}.

The result of a semi-coherent wide-parameter space search is a list of any candidates which
pass a predefined detection threshold. To further vet these candidates,
they are subjected to a \emph{hierarchical follow-up}: a process of increasing the
coherence time, eventually aiming to calculate a fully coherent detection
statistic over the maximal span of data. In essence, the semi-coherent search is
powerful in detecting unknown signals as it spreads the significance of a
candidate over a wider area of parameter space allowing for a sparser
template covering. Subsequently, the follow-up attempts to
reverse this process and recover the maximum significance and tightly constrain
the candidate parameters (cf.\ Section~\ref{sec_follow_up} for an illustration
and introduction to other hierarchical follow-up methods). In this work,
we propose a hierarchical follow-up procedure using \ac{MCMC} optimization.

\ac{MCMC} optimization has been used in
parameter estimation for gravitational waves from compact binary coalescence events
\citep{christensenmeyer1998, christensenmeyer2001,
christensenmeyer2004, rover2006, veitch2015parameter}. For the problem of
detecting \acp{CW}, \ac{MCMC}-based
methods have been developed for directed searches with unknown frequency
and spin-down \citep{christensendupuis2004, umstatter2004}.
In \citet{veitch2007}, a method was developed and
applied for a directed search for \acp{CW} from Supernova remnant 1987A and
more recently
for \acp{CW} from known
pulsars \citep{abbott2010} (since this time, an improved nested sampling based
approach has instead been used for known pulsars \citep{Pitkin2012,
Pitkin2017nested, abbott2017knownpulsar}).

\ac{MCMC} searches for simulated \ac{CW} signals using these methods converge
to the correct solution in a reasonable amount of time, provided the volume of
parameter space occupied by a signal (a concept we define in
Section~\ref{sec_number_of_templates}) is a significant fraction of the total
prior volume. When this is not the case, convergence becomes problematic;
\citet{umstatter2006} discussed this in the context of frequency and spin-down
uncertainty, but uncertainties in sky location will cause similar issues.

In this work, we provide methods to understand the conditions under which
\ac{MCMC} searches are effective. We then apply this to develop a
follow-up method for candidates from semi-coherent wide-parameter space searches.

We begin in Section~\ref{sec_mcmc_samplers} with a description of
\ac{MCMC} samplers and then discuss how these can be used in \ac{CW} searches
in Section~\ref{sec_MCMC_and_the_F_statistic}. In
Section~\ref{sec_understanding_the_search_space} we introduce methods to
understand the conditions for which an \ac{MCMC} search is suitable before
describing and testing the follow-up procedure in Section~\ref{sec_follow_up}.
Finally, in Section~\ref{sec_computing_cost} we give a timing model to predict
computing costs before concluding in Section~\ref{sec_conclusion}.

The methods introduced in this paper have been implemented in the open source
\textsc{python} package \textsc{pyfstat} \citep{pyfstat}. Source code along
with all examples in this work can be found at
\url{https://gitlab.aei.uni-hannover.de/GregAshton/PyFstat}.

\section{\ac{MCMC} samplers}
\label{sec_mcmc_samplers}

Given some data $\data$ and a hypothesis $\H$ about the generative model which
produced the data, we can infer the posterior distribution for the set of model
parameters $\theta$ from Bayes theorem
\begin{align}
P(\theta| \data, \H) = \frac{P(\data | \theta, \H)P(\theta | \H)}{P(\data| \H)}\,.
\label{eqn_bayes}
\end{align}
In this expression, $P(\data | \theta, \H)$ is the \emph{likelihood} of
observing the data for the generative model at hand, $P(\theta| \H)$ is the
\emph{prior} probability distribution for the model parameters, and $P(\data|
\H)$ is a normalization constant.

For many problems, it is not possible to solve Eq.~\eqref{eqn_bayes}
analytically. \ac{MCMC} algorithms provide a means to instead sample from $P(\theta| \data,
\H)$, the \emph{posterior distribution}, allowing inferences about the
model parameters to be made using these samples. Since the
normalization constant is independent of $\theta$, it is only necessary to
sample from the unnormalized distribution, i.e.,
\begin{align}
P(\theta| \data, \H) \propto P(\data | \theta, \H)P(\theta | \H)\,.
\label{eqn_bayes_prop}
\end{align}

In the rest of this Section, we introduce the details of the \ac{MCMC} sampler
used in this investigation. For a general introduction to \ac{MCMC} samplers,
see, e.g., \citet{mackay2003information} or \citet{gelman2013bayesian}.

\subsection{The {\texttt{ptemcee}} sampler}
\label{sec_ptemcee}

In this work we use the \texttt{ptemcee} MCMC ensemble sampler
\citep{vousden2016, foreman-mackay2013}, an implementation of the
affine-invariant ensemble sampler proposed by \citet{goodman2010}. This choice
addresses a key issue with the use of \ac{MCMC} samplers, the choice of
\emph{proposal distribution}. At each step of the \ac{MCMC} algorithm, the
sampler generates from a proposal distribution a jump to a new point in
parameter space. Usually, this proposal distribution must be ``tuned'' so that
the \ac{MCMC} sampler efficiently walks the parameter space without either
jumping too far, or taking such small steps that it takes a long time to
traverse the peak \citep{mackay2003information, gelman2013bayesian}.

\addtext{
Ensemble samplers use a number of
parallel \emph{walkers}. The proposal jump for each walker is generated
by the \emph{stretch move} (see \citet{goodman2010}): the
new position is determined by the position of a complementary walker from the
ensemble and a scaling variable. The scaling variable is a random variable,
for the particular form used here see \citet{foreman-mackay2013}, with a
single proposal scale parameter. This scale parameter does not require tuning
and can typically be left at its default value.
As such, ensemble samplers do not use a proposal distribution
which requires tuning; in effect the proposal is determined by the position
of the others walkers in the ensemble.}

Moreover, by applying an affine
transformation, the efficiency of the algorithm is not diminished when the
parameter space is highly anisotropic. Aside from the parallel tempering
functionality (which we discuss momentarily), this sampler has
three tuning parameters: a proposal scale, the number of walkers $\Nwalkers$,
and the number
of steps to take. For the number of walkers, it is
recommended \citep{foreman-mackay2013} to use as many as possible (limited by
memory constrains); for our purposes we typically find that $\Nwalkers=100$ is
sufficient. We will discuss the number of steps in
Section~\ref{sec_assessing_convergence}.

When setting up an ensemble \ac{MCMC} sampler, one must also consider how to
\emph{initialize} the sampler, choosing the initial parameter values for each
walker. Typically, we want to explore the entire prior parameter space and therefore
the initial position of each walker can be selected by a random draw from the
prior distribution. However, instances do occur when one would like to
initialize the walkers from a different distribution, e.g., if a search
has already been performed and the signal localized to a small region of the prior
volume and one would solely like to perform parameter estimation.

\subsection{Parallel tempering: sampling multimodal posteriors}
\label{sec_parallel_tempering}
Beyond the standard ensemble sampler, the \texttt{ptemcee} sampler also uses
\emph{parallel-tempering}. A parallel
tempered \ac{MCMC} sampler \citep{swendsen1986} runs
$\Ntemps$ simulations in parallel with the likelihood in the $i\mathrm{th}$
simulation raised to a power $1/\temp_i$, where $\temp_i$ is referred
to as the \emph{temperature}. Eq.~\eqref{eqn_bayes_prop} for
the $i$th temperature is then rewritten as
\begin{equation}
P(\theta | \data, \temp_i, \H) \propto P(\data| \theta, \H)^{1/\temp_i}P(\theta| \H)\,.
\end{equation}
Setting $\temp_0=1$ with $\temp_i > \temp_0 \; \forall \; i > 0$, such that the $i=0$
temperature recovers Eq.~\eqref{eqn_bayes_prop} while for higher
temperatures the likelihood is broadened. During the
simulation, the algorithm swaps the position of the walkers between the
different temperatures. This allows the $\temp_0$ walkers (from which we draw samples
of the posterior) to efficiently sample from a multimodal posterior. This
method introduces additional tuning parameters for the choice of temperatures.
\addtext{Unless stated otherwise, all examples in this work use a default setup:
\CHECK{3} temperatures
logarithmically spaced between $\CHECK{1}$ and $\CHECK{10^{0.5}}$. In
testing (see for example Section~\ref{sec_maximim_size_of_prior})
this was found to be efficient over a range of searches}. \texttt{ptemcee}
also implements a dynamical temperature selection algorithm \citep{vousden2016}
which will update the temperature ladder during the sampling.

\subsection{Assessing convergence and independent samples}
\label{sec_assessing_convergence}

In tuning the number of steps taken by a Markov chain there are two distinct
issues to address: initialization bias and autocorrelation in equilibrium
\citep{sokal1997}.

i) The first of these issues refers to the initial transient period which
occurs whilst the Markov chain (in our case, the \ac{MCMC} sampler) transitions
from the initialization distribution to the posterior distribution. To remove any bias, samples
taken during this period, referred to as the \emph{burn-in}, are discarded from
the set of samples used to infer the posterior. An example
of this burn-in period can be seen later in Fig.~\ref{fig_MCMC_simple_example}. The
difficulty lies in deciding how many burn-in steps are required, values of
$\mathcal{O}(10^2-10^4)$ steps are typical in the literature. In most cases,
$\Nburn$, the number of burn-in steps, is predetermined and a graphical check of
the \ac{MCMC} samples can be performed to ensure the validity of the results.
To determine the number of burn-in steps needed, \citet{sokal1997} defines the
\emph{exponential autocorrelation time} $\tau_\mathrm{exp}$, the relaxation
time of the slowest mode of the system, and recommends that
``$20\tau_\mathrm{exp}$ burn-in samples are usually more than adequate''.

ii) The second issue, is that once the sampler has reached
equilibrium it will draw samples from the posterior distribution (we refer to
these are the \emph{production} samples), but these samples are not
independent, they are correlated with an \emph{integrated autocorrelation time}
$\tau_\textrm{int}$ \citep{sokal1997}.  The effective number of independent
samples from a simulation with $\Nprod$ production samples is then roughly
$\Nprod/2\tau_\mathrm{int}$.

The total number of steps taken by the sampler is the
sum of $\Nburn$ and $\Nprod$.

Usually, $\tau_\mathrm{exp}$ and $\tau_\mathrm{int}$ are of the same order of
magnitude \citep{sokal1997}. \meta{Therefore, we will use the \texttt{emcee}
estimator for the integrated autocorrelation (see \citet{foreman-mackay2013})
time which we refer to simply as
the \ac{ACT}}. In future work, we plan to explore the method proposed by
\citet{akeret2013} and \citet{allison2014} to estimate $\tau_\mathrm{exp}$ from
a least-squares fit to the autocorrelation function and also to investigate
stopping criteria. In Section~\ref{sec_example} and
Section~\ref{sec_computing_cost}, we will provide
estimates for the \ac{ACT} in the context of \ac{CW} searches.

\meta{The methods discussed here are part of a broader literature on assessing
convergence. We have chosen to use the autocorrelation time following the
recomendations given in \citet{foreman-mackay2013}, but alternatives such as
the method proposed by \citet{raftery1991many} could also be applied.  For a
broad review and general discussion of convergence, see
\citet{cowles1996markov} and \citet{hogg2017}.}

\section{\ac{MCMC} and the $\F$-statistic}
\label{sec_MCMC_and_the_F_statistic}

In this section, we give a brief introduction to the application of the
$\F$-statistic in a hypothesis testing framework and then show how this
can be used in a \ac{CW} \ac{MCMC} search.

\subsection{Hypothesis testing framework: fully coherent}
\label{sec_hypothesis_testing}

In \citet{prix2009}, a framework was introduced demonstrating the use of the
$\F$-statistic \citep{jks1998, cutlershutz2005} in defining the \emph{Bayes
factor},
\begin{equation}
\Bsg(\data) \equiv \frac{P(\data| \Hs)}{P(\data| \Hg)}\,,
\label{eqn_Bsg_x}
\end{equation}
between the signal hypothesis and the Gaussian noise hypothesis computed on
some data $\data$; we now briefly review this framework.

\meta{
The noise hypothesis $\Hg$ states that the data $\data$ contains
only Gaussian noise $\noise$, while the signal hypothesis $\Hs$
states that there is an additive signal in the data, i.e.\
$\data = \noise + \signal(\A,\blambda)$. As mentioned in the
introduction, the signal $\signal$ is generally expressed as
depending on two sets of parameters, namely the four
amplitude parameters $\A$, and a number of
phase-evolution parameters $\blambda$, such as the frequency,
spindown and sky-position of the source.
}

\meta{
With this one obtains an explicit analytic expression (not given here
for brevity, e.g.\ see \cite{prix2009gravitational}) for the
\emph{likelihood-ratio function}, defined as
}
\begin{equation}
\mathcal{L}(\data; \A, \blambda) \equiv
\frac{P(\data| \Hs, \A, \blambda)}
{P(\data| \Hg)}\,.
\end{equation}
\meta{
Assuming independent priors for the amplitude parameters $\A$ and the
phase-evolution parameters $\blambda$, and noting that $P(\data| \Hg)$
does not depend on these signal parameters, we can express the Bayes
factor of Eq.~\eqref{eqn_Bsg_x} as
}
\begin{equation}
\Bsg(\data) = \iint
\mathcal{L}(\data; \A, \blambda) P(\A| \Hs)P(\blambda| \Hs) \; d\A d\blambda\,.
\label{eqn_Bsg_x_A_lambda}
\end{equation}
It was shown by \citet{prix2009} \meta{(see also
\citet{2014CQGra..31f5002W} for a more detailed discussion)} that with an
appropriate choice of
$P(\A|\Hs)$, the marginalization over the amplitude parameters
\begin{equation}
\Bsg(x; \blambda) = \int
\mathcal{L}(\data; \A, \blambda) P(\A| \Hs)\; d\A\,,
\end{equation}
can be performed analytically. We refer to the Bayes factor
$\Bsg(x;\blambda)$ for a fixed value of the phase-evolution parameters
$\blambda$ as a ``targeted'' Bayes factor. Once $\Bsg(\data; \blambda)$ is computed,
the full Bayes factor of Eq.~\eqref{eqn_Bsg_x_A_lambda} can be calculated by numerical integration
over the phase-evolution parameters, i.e.\
\begin{equation}
\Bsg(\data) = \int \Bsg(\data; \blambda)P(\blambda| \Hs)\; d\blambda\,.
\label{eqn_Bsg_x_lambda}
\end{equation}
In this work, we use the amplitude prior choice suggested in \citet{prix2011},
which results in
\begin{equation}
\Bsg(\data; \blambda) \equiv \frac{70}{\rhohatmax^{4}}
e^{\Ftilde(\data; \blambda)}\,,
\label{eqn_targeted_bayes_factor}
\end{equation}
where $\Ftilde$ is the fully coherent $\F$-statistic.
\meta{
This statistic was originally found \citep{jks1998,cutlershutz2005} as
a frequentist amplitude-maximized likelihood-ratio statistic, i.e.\
\begin{equation}
  \label{eq:1}
  \Ftilde(\data;\blambda) \equiv \max_{\A} \ln\mathcal{L}(\data;\A,\blambda)\,,
\end{equation}
which can be expressed in closed analytic form for any given $\blambda$.
The $\F$-statistic follows a (non-central) $\chi^2_4$ distribution
with four degrees of freedom, with expectation value
\begin{equation}
  \label{eq:2}
  E\left[2\Ftilde\right] = 4 + \rhotilde^2\,,
\end{equation}
where $\rhotilde$ defines the coherent \ac{SNR}, and
$\rhotilde^2$ is the non-centrality parameter of the $\chi^2$ distribution.
}

We note that the ``modified ad-hoc $\F$-statistic prior'' of \citet{prix2011} leading
to Eq.~\eqref{eqn_targeted_bayes_factor} is somewhat
unphysical in that it introduces an arbitrary cutoff $\rhohatmax$ for the
signal strength and is not uniform in $\cos\iota$.
One could circumvent this by numerically integrating
Eq.~\eqref{eqn_Bsg_x_lambda} with physical amplitude priors.
However, by using the $\F$-statistic instead, we can use
fast and mature codes implementing this statistic (namely \textsc{XLALComputeFstat()}
\citep{lalsuite}) and thereby greatly improving the speed at which a
search can be run with very little loss of detection power \citep{prix2009}.

The \ac{MCMC} class of optimization tools are formulated to solve the problem of
inferring the posterior distribution for some general model parameters given
some data~$\data$.  In this case, we
are concerned with the posterior distribution for $\blambda$, the
phase-evolution parameters which cannot be marginalized analytically. In
analogy with Eq.~\eqref{eqn_bayes_prop}, this distribution is given by
\begin{equation}
P(\blambda| \data, \Hs) \propto P(\data | \blambda, \Hs) P(\blambda| \Hs)\,.
\end{equation}
Dividing through by $P(\data| \Hg)$, we can then write this as
\begin{equation}
P(\blambda | \data, \Hs) \propto \Bsg(\data; \blambda)P(\blambda | \Hs)\,.
\label{eqn_lambda_posterior}
\end{equation}

We use an MCMC algorithm to perform a CW search by applying
Eq.~\eqref{eqn_targeted_bayes_factor} as the likelihood along with a prior for
the phase-evolution parameters. In this way, the log-likelihood function is
proportional to $\Ftilde$.

\subsection{Hypothesis testing framework: semi-coherent}

Eq.~\eqref{eqn_lambda_posterior} is the posterior distribution using a fully coherent
statistic (i.e., the log-likelihood is given by $\Ftilde$). It was shown
by \citet{prix2011}, that the semi-coherent $\F$-statistic naturally arises by
splitting the likelihood into $\Nseg$ independent non-overlapping
\emph{segments} and allowing for independent choices of the amplitude
parameters in each segment. Specifically, the semi-coherent targeted Bayes factor is
\begin{equation}
\Bsg(\data;\blambda, \Nseg) \equiv
\left(\frac{70}{\rhohatmax^{4}}\right)^{\Nseg}
e^{\Fhat}\,,
\end{equation}
where the semi-coherent\footnote{We identify semi-coherent quantities by a ``hat''
and fully coherent quantities with a ``tilde''.}  $\F$-statistic is
\begin{equation}
\Fhat \equiv \sum_{\ell=1}^{\Nseg}\Ftilde(\data_{(\ell)}; \blambda)\,,
\label{eqn_sc_targeted_bayes_factor}
\end{equation}
and $\data_{(\ell)}$ refers to the data in the $\ell\mathrm{th}$
segment.
\meta{
This semi-coherent $\F$-statistic follows again a (non-central)
$\chi^2$-distribution, but with $4\,\Nseg$ degrees of freedom, and
non-centrality parameter
\begin{equation}
  \label{eq:3}
  \rhohat^2 = \sum_{\ell=1}^{\Nseg} \rhotilde_\ell^2\,,
\end{equation}
with per-segment SNRs $\rhotilde_\ell$, and expectation
\begin{equation}
  \label{eq:5}
  E\left[2\Fhat\right] = 4\,\Nseg+\rhohat^2\,.
\end{equation}
}

The posterior distribution for $\blambda$, Eq.~\eqref{eqn_lambda_posterior},
naturally generalizes using a semi-coherent statistic to
\begin{equation}
P(\blambda | \data, \Nseg, \Hs) \propto \Bsg(\data; \blambda, \Nseg)P(\blambda | \Hs)\,.
\label{eqn_lambda_posterior_sc}
\end{equation}
and hence can equally be applied in an MCMC algorithm.

One can consider the fully coherent case as a special instance of the
semi-coherent case with $\Nseg=1$. Any methods developed for parameter
estimation using the fully coherent $\F$-statistic can easily be generalized to
the semi-coherent case. Therefore, in the remainder of this Section, we
discuss the general intricacies of setting up and running an \ac{MCMC} search
using the simpler fully coherent $\F$-statistic.  In
Section~\ref{sec_follow_up}, we turn to the primary goal of this paper,
the hierarchical multi-stage follow-up of candidates found in semi-coherent
searches.

\subsection{Example: fully coherent \ac{MCMC} optimization for a signal in noise}
\label{sec_example}

In order to familiarize the reader with the features of an \ac{MCMC}-CW search, we
will now describe a simple directed search (over $f$ and $\dot{f}$) for a
simulated signal in \CHECK{\BasicExampleDays}~days of Gaussian-noise data. The signal is generated
with an amplitude $\CHECK{\BasicExamplehzero}$ while for the Gaussian noise
$\sqrt{\Sn}=\CHECK{\BasicExampleSqrtSn}$~\noiseUnits (at the fiducial frequency
of the signal), such that the signal has a
\emph{sensitivity depth} of
\begin{equation}
  \label{eq:6}
  \Depth\equiv\frac{\sqrt{\Sn}}{h_0} = \CHECK{\BasicExampleDepth}\,\mathrm{Hz}^{-1/2}\,.
\end{equation}
The remaining signal parameters are chosen
randomly and labeled $\blambda\Signal$ and $\A\Signal$ such that the simulated
frequency and spin-down values are $f\Signal$ and $\fdot\Signal$; we also choose the
reference time to coincide with the middle of the data span.

First, we must define a prior for each search parameter. Typically, we use
either a uniform prior bounding the area of interest, but a more informative
distribution centered on the target with some uncertainty could also be used.
For this example, we use a uniform prior with a frequency range of $\Delta
f=\CHECK{\BasicExampleDeltaFzero}$~Hz and a spin-down range of $\Delta
\fdot=\CHECK{\BasicExampleDeltaFone}$~Hz/s both centered on the simulated
signal frequency and spin-down rate. These values were chosen such that the
approximate number of unit-mismatch templates is
$\N^*\approx\CHECK{\BasicExampleNstar}$ (defined in
Section~\ref{sec_number_of_templates}), which, as we discuss in
Section~\ref{sec_maximim_size_of_prior}, is sufficiently small to
ensure the MCMC search is effective.

Having defined the prior, we initialize the positions of the
$\CHECK{\BasicExamplenwalkers}$ walkers randomly from the prior. The final
setup step is to define the number of burn-in and production steps the sampler
should take; this is a tuning parameter of the \ac{MCMC} algorithm, but should
be informed by the maximum \ac{ACT} (see Sec.~\ref{sec_assessing_convergence}).
For this example, we find this to be $\sim \CHECK{30}$, so to ensure the
burn-in length is sufficient $\Nburn \gtrsim \CHECK{300}$ (i.e., \CHECK{10}
times the \ac{ACT}), while to generate $\sim\CHECK{1000}$ independent samples, we
need each walker to perform $\Nprod=\CHECK{300}$ steps.

\begin{figure}[htb]
\centering
\includegraphics[]{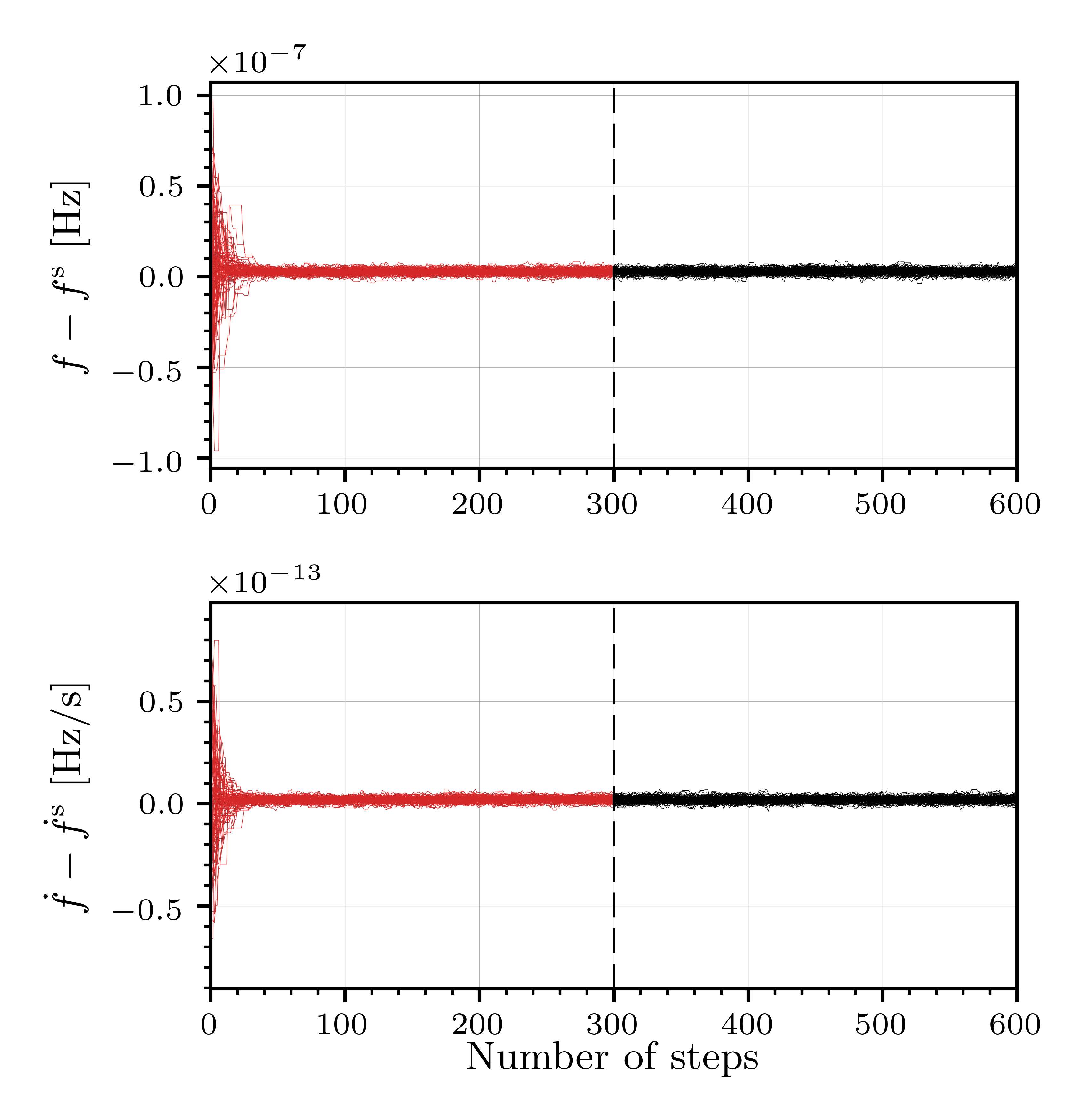}
\caption{
The progress of each walker in the \ac{MCMC} simulation for a simulated signal in Gaussian
noise, searching over a \emph{small} directed search parameter space:
the approximate number of unit-mismatch templates, defined in
Section~\ref{sec_number_of_templates}, is $\N^*\approx \CHECK{\BasicExampleNstar}$.}
\label{fig_MCMC_simple_example}
\end{figure}

Using these choices, the simulation is run. To illustrate the full \ac{MCMC}
process, in Fig.~\ref{fig_MCMC_simple_example} we plot the progress of all
100 individual walkers (each represented by an individual line) as a function
of the total number of steps. The portion of steps to the left of the dashed
vertical line are burn-in samples (see Section~\ref{sec_assessing_convergence})
and hence discarded. From this plot we see why: the walkers are initialized
from the uniform prior and initially spend some time exploring the whole
parameter space before converging. The production samples from the
converged region, those to the right of the
dashed vertical line, can be used to generate summary statistics or
posterior plots.

\section{Understanding the search space}
\label{sec_understanding_the_search_space}

In general, \ac{MCMC} samplers are highly effective in generating samples of the
posterior in multi-dimensional parameter spaces. However, they will perform
poorly if the volume occupied by the signal is small compared to the prior
volume; in this Section we define exactly what is meant by this for CW
searches.

\subsection{The metric}
\label{sec_metric}

In a traditional \ac{CW} search that uses a template bank, the spacing of
the grid are chosen such that the loss of relative \ac{SNR} is bounded. The
optimal choice of grid then consists of minimizing the computing cost while
respecting this bound \citep{pletsch2010, prix2012, wette2013, wette2015}. We
now discuss how the work on setting up these
grids can be applied to the problem of determining whether the setup is
appropriate for an \ac{MCMC} method.

For an $\F$-statistic search on data containing
a signal with phase-evolution parameters $\blambda\Signal$, we define the
\emph{mismatch} $\mu$ at a template point $\blambda$ as
\begin{align}
\mu(\blambda\Signal, \blambda) \equiv
\frac{\rho^2(\blambda\Signal; \blambda\Signal) - \rho^2(\blambda\Signal; \blambda)}
{\rho^2(\blambda\Signal; \blambda\Signal)}
  \in [0, 1]\,,
\label{eq:4}
\end{align}
\meta{
where $\rho^2(\blambda\Signal ; \blambda)$ is the non-centrality
parameter at $\blambda$, given
a signal at $\blambda\Signal$; as such $\rho^2(\blambda\Signal;
\blambda\Signal)$ is the perfectly-matched non-centrality parameter,
which is generally proportional to the squared \ac{SNR} (with equality
only in the fully coherent case of $\Nseg=1$).
}

For small offsets $\Delta\blambda \equiv \blambda - \blambda\Signal$ between the template
and signal, the mismatch of Eq.~\eqref{eq:4} can be approximated by
Taylor-expanding in small $\Delta\blambda$, defining the \emph{metric
  mismatch} \citep{owen1996search, balasubramanian1996,brady1998} as
\begin{equation}
\mu(\blambda, \Delta\blambda) \equiv
g_{ij} \Delta\lambda^{i}\Delta\lambda^{j}
+ \mathcal{O}\left(\Delta\blambda^{3}\right)\,,
\label{eqn_mutilde_expansion}
\end{equation}
where $g_{ij}$ is referred to as the \emph{metric} and provides a measure of
distances in parameter space. Generally, Eq.~\eqref{eqn_mutilde_expansion}
is a good approximation up to $\mu \gtrsim 0.3-0.5$ \citep{prix2007, wette2013}.

In general, no analytic approximation of the full phase-metric exists, although
analytic solutions can be found when searching at a single point in sky for the
frequency and spin-down components. A numerical approximation was formulated by
\citet{wette2013} for the fully coherent phase-metric which uses a coordinate
transformation, to the so-called \emph{reduced supersky metric}
$\tildegrss_{ij}$, which, of significance in the next section, is
well-conditioned.  In \citet{wette2015}, the work was further generalized to
the semi-coherent case $\hatgrss_{ij}$ when there are $\Nseg$ segments; in the
following we use this later definition, with the understanding that when
$\Nseg=1$, $\hatgrss_{ij} = \tildegrss_{ij}$.

\subsection{The number of templates}
\label{sec_number_of_templates}

When constructing a lattice of search templates (a template bank) to cover an
$n$-dimensional parameter space $\mathbb{P}$ at a fixed maximal mismatch of
$\mu_0$, it can be shown \citep{prix2007, messenger2009, brady1998,
owen1996search, owen1999matched} that the required
number of templates is given by
\begin{align}
\N = \theta \mu_0^{-n'/2} \int_{\mathbb{T}_{n'}}
\sqrt{\textrm{det}g} \,d^{n'}\lambda\,,
\label{eqn_Nnprime}
\end{align}
where $g_{ij}$ the parameter space metric, $\mathbb{T}_{n'} \subseteq \mathbb{P}$ is
the $n'\le n$ dimensional space spanned by the template bank, and $\theta$ is the normalized
thickness, which depends on the geometric structure of the covering.

A way to understand the ``size'' of a given parameter space compared to the
size a signal would occupy is to define
\begin{align}
\Neff_{n'} \equiv \int_{\mathbb{T}_{n'}}\sqrt{\textrm{det}g} \,d^{n'}\lambda\,,
\label{eqn_Nstar}
\end{align}
the approximate number of templates needed to cover the parameter space at a
mismatch of unity. It is the approximate number in the sense that we neglect
the effects of the normalized thickness, assuming that $\theta=1$. In general,
$\theta$ will depend on the geometric structure of a given covering and the
number of dimensions (see, e.g., \citet{prix2007} or \citet{messenger2009}).
Nevertheless, we fix $\theta=1$
in order to give a rough order of magnitude estimate.

A subtle point \citep{brady2000, cutler2005, prix2012} is that
$n'$, the number of dimensions of the template bank space $\mathbb{T}_{n'}$
may be less than $n$, the number of dimensions of the search parameter space
$\mathbb{P}$. This is because a given dimension of the search space may be
under-resolved and hence require only one signal template. When
calculating the approximate number of templates, only those dimensions which
are fully resolved should be included.  Pragmatically, as done in
\citep{brady1998, brady2000, cutler2005, prix2012},
this can be calculated as
\begin{equation}
\Neff = \max_{n'}\Neff_{n'}\,,
\label{eqn_N}
\end{equation}
which we refer to as the \emph{approximate number of unit-mismatch templates}
needed to cover a given search parameter space. We use this to quantify the
size of the parameter space.

For follow-up searches, typically the size of our uncertainties are small
compared to the scale of parameter space correlations, in other words the
metric can be assumed to be constant over that space. As such, Eq.~\eqref{eqn_Nstar} can be
approximated as
\begin{equation}
\Neff_{n'} \approx \sqrt{\textrm{det}g}\; \mathrm{Vol}(\mathbb{T}_{n'})\,,
\end{equation}
where $\mathrm{Vol}(\mathbb{T}_{n'})$ is the coordinate-volume of the
template-bank parameter space.

In the context of an \ac{MCMC} search, for a uniform prior, the search parameter
space is a hypercube and so $\mathrm{Vol}(\mathbb{T}_{n'})$ can be calculated
from the product of the
edges. However, the metric in the usual coordinate systems is often
ill-conditioned and for dimensions above 3 it becomes difficult numerically to
calculate the determinant. To circumvent this issue, we compute the volume
in the reduced supersky coordinates for which the metric is well-conditioned
\citep{wette2013}. In this coordinate
system, the search parameter space will in general be a parallelotope for which
the volume can be computed after appropriate transformation of the search
parameter space hypercube. The numerical computation of the reduced supersky
metric and associated coordinate transformations are handled by
\textsc{ComputeSuperskyMetrics()}, a routine in \textsc{LALPulsar}
\citep{lalsuite}.

For an \ac{MCMC} search, its possible to initially restrict the search to a
subset of the prior volume by appropriate initialization of the walkers. In
this instance, $\N^*$ should be calculated using the volume of this subset, and
not the prior volume.

\subsection{The topology of the likelihood}
\label{sec_topology}

We use the $\F$-statistic as a log-likelihood in our \ac{MCMC} searches.
To understand how such an MCMC search may behave, it is worthwhile to acquaint
ourselves with the typical topology of the
$\twoFtilde$ surface, i.e., how the detection statistic varies over the unknown
parameters in the presence of noise, or a signal and noise.

\begin{figure}[tb]
\centering \includegraphics[width=0.5\textwidth]{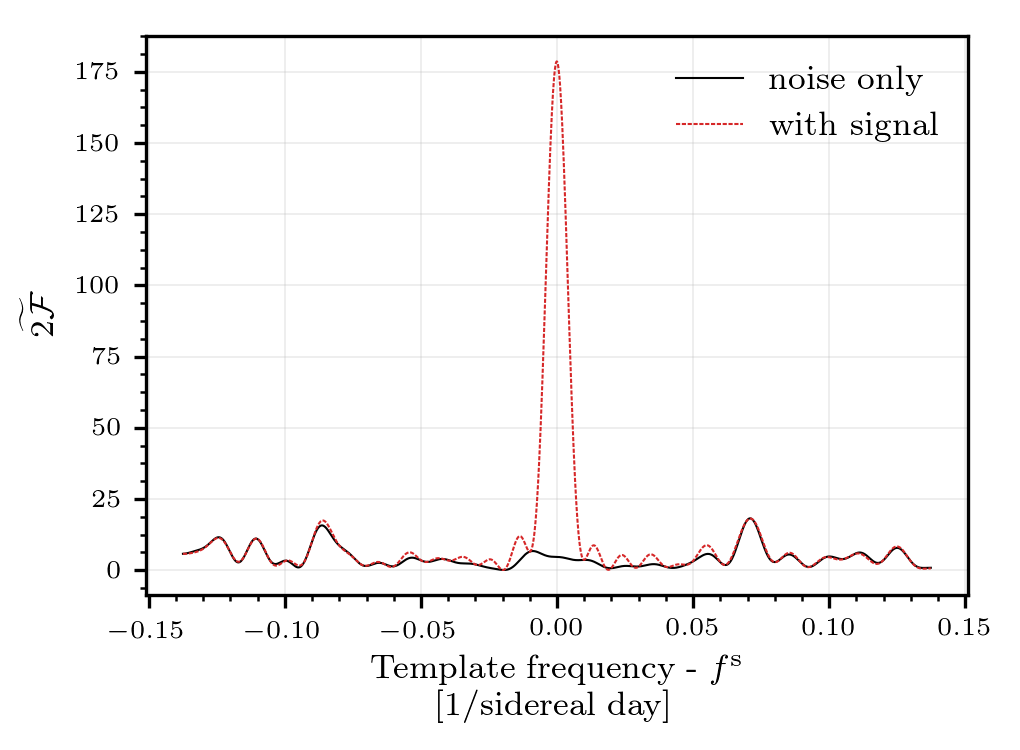}
\caption{Numerically computed $\twoFtilde$ for data (lasting
\GridedFrequencyInitialT~days) containing Gaussian noise only and the same data with
an added simulated signal at frequency $f\Signal$.}
\label{fig_grid_frequency}
\end{figure}

The expectation of $\twoFtilde$ is $4$ in Gaussian noise alone, but
in the presence of a signal is $4+\rho^{2}$ \citep{jks1998, cutlershutz2005}
where $\rho^{2}$ is the non-centrality parameter, see
Eqs.~\eqref{eq:2} and \eqref{eq:5}.
To illustrate the behavior in noise alone, in Fig.~\ref{fig_grid_frequency} we
plot $\twoFtilde$ over the template frequency for a random instance of Gaussian
detector noise with
$\sqrt{\Sn}=$\CHECK{\GridedFrequencyInitialSqrtSx}~\noiseUnits~lasting for
\CHECK{\GridedFrequencyInitialT}~days: we see fluctuations with multiple
maxima.

Taking the same instance of Gaussian noise, we add to the data a simulated signal
($h_0=$\CHECK{\GridedFrequencyInitialhzero}) with a frequency $f\Signal$ (other parameters are
chosen arbitrarily and not critical for this illustrative example).
In Fig.~\ref{fig_grid_frequency}, in the presence of a signal, the $\twoFtilde$
surface peaks close to the simulated signal frequency.  In addition to this
central peak, secondary signal-maxima can be seen; indeed, it can be shown
analytically (i.e., using Eq.~(11) of \citet{prix2005} with $\Delta\Phi(t)
= 2\pi t(f - f\Signal)$) that in the presence of a signal, $\twoFtilde$ follows a
$\mathrm{sinc}^{2}$ behavior.  Away
from the signal peak, the $\twoFtilde$ surface approximately agrees with the
noise-only result as expected.

\begin{figure}[htb]
\centering \includegraphics[width=0.5\textwidth]{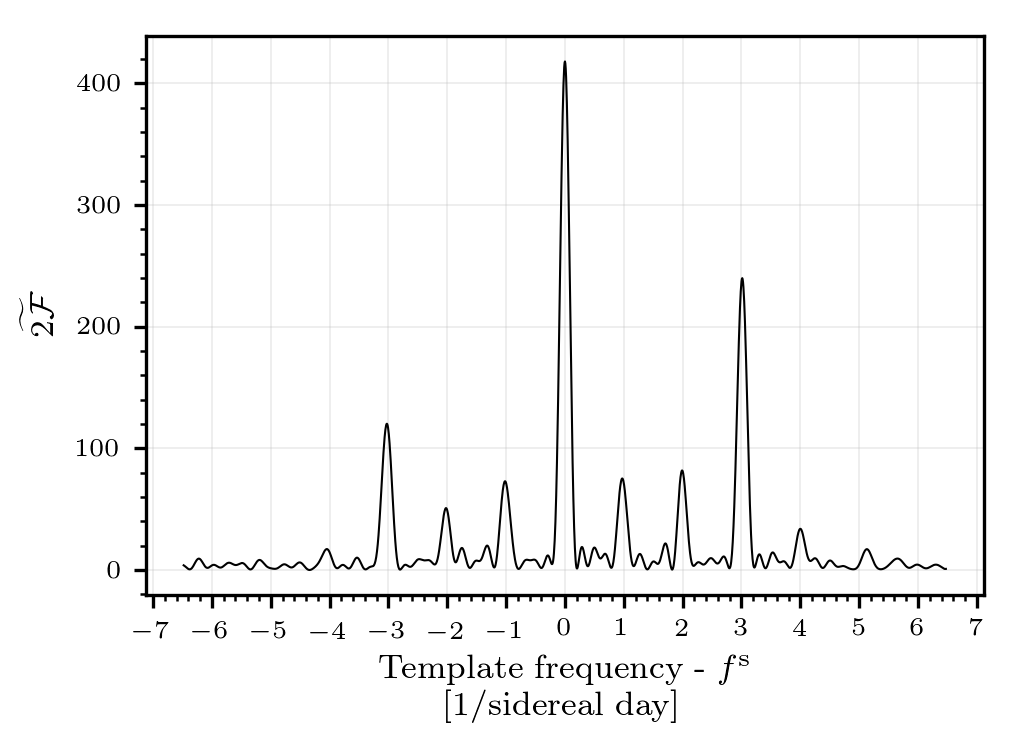}
\caption{Numerically computed $\twoFtilde$ for a
simulated signal with frequency $f\Signal$ in Gaussian noise lasting for $\GridedFrequencyT$.
Additionally, the signal was simulated with $\delta=\GridedFrequencyDelta$~rad
and $\cos(\iota)=\GridedFrequencycosi$.}
\label{fig_grid_frequency_short}
\end{figure}

In addition to the $\mathrm{sinc}^{2}$ response for a signal, there are
additional \emph{sideband} peaks spaced at intervals of the inverse sidereal
day; the presence of these can be deduced from Eq.~(12) and (13) of
\citet{jks1998}, but for a detailed discussion see
\citet{sammut2014implementation}. To demonstrate this, in
Fig.~\ref{fig_grid_frequency_short} we plot the $\twoFtilde$ surface over the
frequency for a signal lasting \GridedFrequencyT~days, such that the sideband
spacing is approximately an order of magnitude larger than the signal width.

Here, there are 9 peaks (8 of which are distinctive in this figure)
spaced in intervals of the inverse sidereal
day (the central peak being at the simulated signal frequency). The relative
prominence of each side-peak varies strongly as a function of the declination
in the sky and $\iota$, the angle between the spin-axis of the neutron star and
the line of sight.

In Fig.~\ref{fig_grid_frequency} and \ref{fig_grid_frequency_short}, we have
demonstrated that a signal in noise can have a complicated topology in the
frequency space alone: having a main peak with associated $\mathrm{sinc}^{2}$
structure, sidebands, and other peaks due to Gaussian noise. In practice, we
simultaneously search over other dimensions such as derivatives of the
frequency and sky location which will also have a complicated structure with
multiple peaks.

When running an \ac{MCMC} simulation we must therefore be aware that in
addition to the signal peak, the likelihood will contain multiple modes which
may be either noise fluctuations, secondary peaks of the signal, or the signal
peak itself. It is for these reasons that the parallel-tempered MCMC sampler
(see Sec.~\ref{sec_parallel_tempering}) is applied which can efficiently sample
from such a multimodal posterior.

\subsection{Maximum size of prior parameter space}
\label{sec_maximim_size_of_prior}

For a fixed sized search parameter space, any optimization routine, given
finite resources, will fail if this space is too large. We quantify the size of
a given space by $\N^*$, the approximate number of unit-mismatch templates.  We now
investigate the maximum $\N^*$ that this \ac{MCMC} search can efficiently explore
and explain what happens beyond this point.

\begin{figure}
\centering
\includegraphics[width=0.45\textwidth]{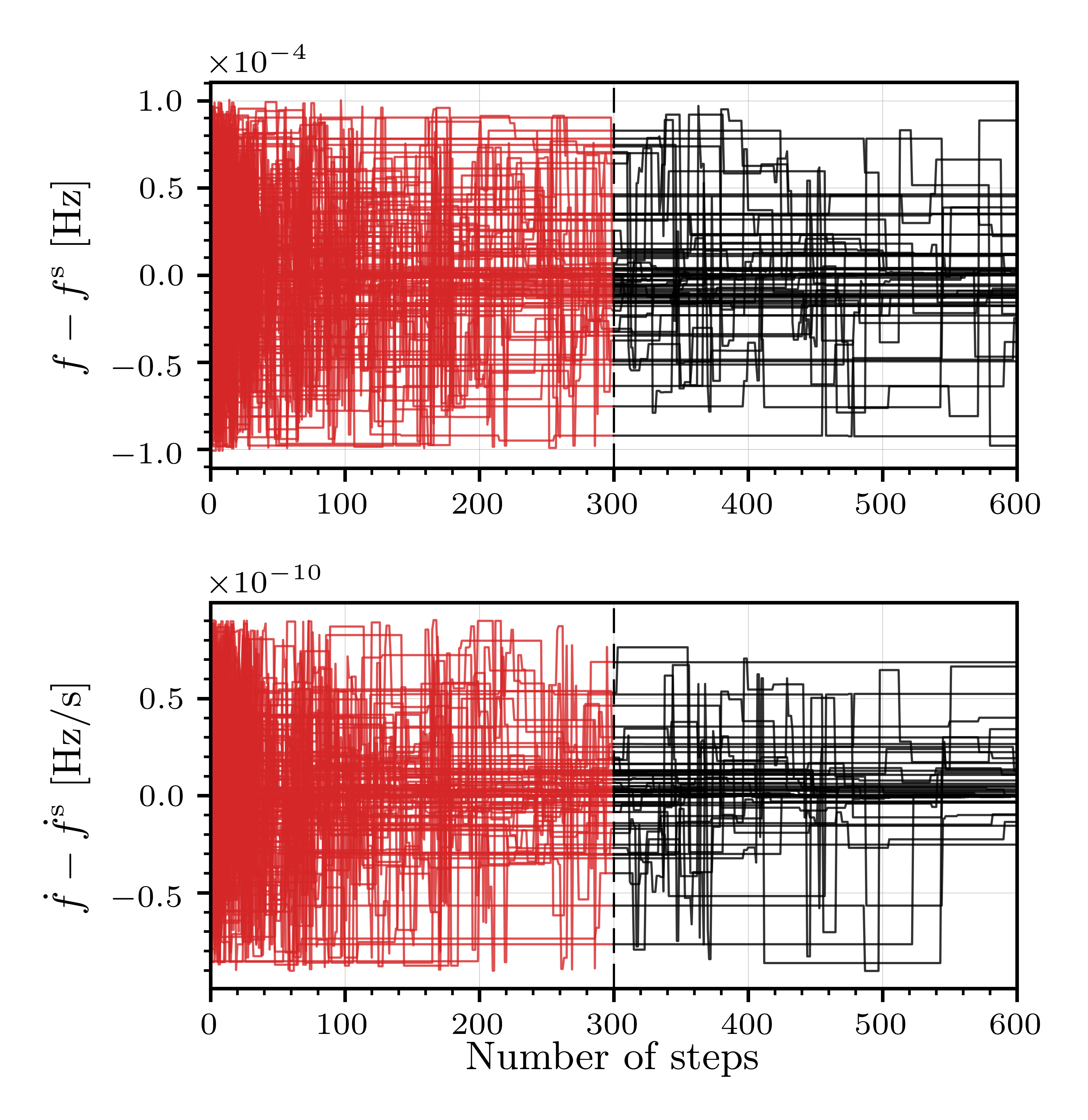}
\caption{
The progress of each walker in the \ac{MCMC} simulation for a simulated signal in Gaussian
noise, searching over a \emph{large} directed search parameter space:
the approximate number of unit-mismatch templates, defined in
Section~\ref{sec_number_of_templates}, is $\N^*\approx \CHECK{\BasicExampleFailNstar}$.}
\label{fig_MCMC_non_convergence}
\end{figure}

An \ac{MCMC} search is efficient if the walkers converge to the global
maxima of the search space, an example can be seen in
Fig.~\ref{fig_MCMC_simple_example}. On the other hand, if the search parameter
space is too large compared to the size that the signal occupies, the walkers
do not converge. Typically, this results in the walkers sampling the noisy
background. In such cases, the search is, at best, an approximation of a
random template bank \citep{messenger2009},
but more often we find that the walkers end up getting stuck in multiple
isolated maxima. As an example, in Fig.~\ref{fig_MCMC_non_convergence},
we repeat the directed search shown in Fig.~\ref{fig_MCMC_simple_example},
but with a factor $\CHECK{10^3}$ larger prior width in both $\Delta f$ and
$\Delta \dot{f}$ such that the approximate number of templates is
$\N^*\approx\CHECK{\BasicExampleFailNstar}$. Clearly, the MCMC search does not
converge.

To quantify whether or not a simulation has converged, let us consider the resulting
set of samples
$\theta^{i}_{m}$, where $\theta\subseteq \blambda$, the upper index $i\in[1, N]$
labels the sample number, while the lower index $m\in[1, M]$ labels the walker
number (recalling that we use an ensemble sampler). If we define the per-walker
mean and variance as
$\hat{\theta}_m$ and $\sigma_m^2$ and the mean over all walkers and steps to be
$\hat{\theta}$, then the between-walker
variance and the within-walker variance are given by
\begin{align}
B_\theta &\equiv \frac{N}{M-1}\sum_{m=1}^{M}(\hat{\theta}_m - \hat{\theta})^2
\end{align}
and
\begin{align}
W_\theta &\equiv \frac{1}{M}\sum_{m=1}^{M}(\hat{\theta}_m)^2\,.
\end{align}
We can then define an intuitive way to quantify if the walkers have converged
by calculating the ratio
\begin{align}
\mathcal{Q}_\theta&\equiv\frac{B_{\theta}}{W_\theta}\,.
\label{eqn_Q}
\end{align}
for $\mathcal{Q}_\theta \gg 1$, the within-walker variance is much smaller
than the variance between walkers, indicative that they are not converged, as
in Fig.~\ref{fig_MCMC_non_convergence}, while
$\mathcal{Q}\sim 1$ indicates they have converged, as in
Fig.~\ref{fig_MCMC_simple_example}.

This ratio is closely related to the Gelman-Rubin statistic \citep{gelman1992,
brooks1998} which was developed to monitor convergence of single-chain samplers
by running multiple independent \ac{MCMC} simulations. There are however theoretical
issues in using the Gelman-Rubin statistic for ensemble samplers that go beyond
the scope of this paper. For now, we find it is sufficient to use the intuitive ratio
defined in Eq.~\eqref{eqn_Q} to assess safe values of $\N^*$. A possible
alternative might be to use the \ac{ACT}. However, to safely
calculate this (using the methods defined in \citet{vousden2016, foreman-mackay2013}), one
needs to run the simulation until convergence. Since we want to investigate
when (in a reasonable amount of computation time) the walkers do not converge,
it is simpler to use $\mathcal{Q}_{\theta}$.

In a data set of \CHECK{10} days, we simulate Gaussian noise and a signal with
a sensitivity depth (defined in Eq.~\eqref{eq:6}) of $\Depth=\CHECK{40}$. Then, we perform a directed search
(over $f$ and $\fdot$). The prior for the search is uniform in frequency and
spin-down centered on the simulated signal and chosen such that the effective
number of templates is $\N^*$.

\begin{figure}[htb]
\centering
\includegraphics[width=0.49\textwidth]{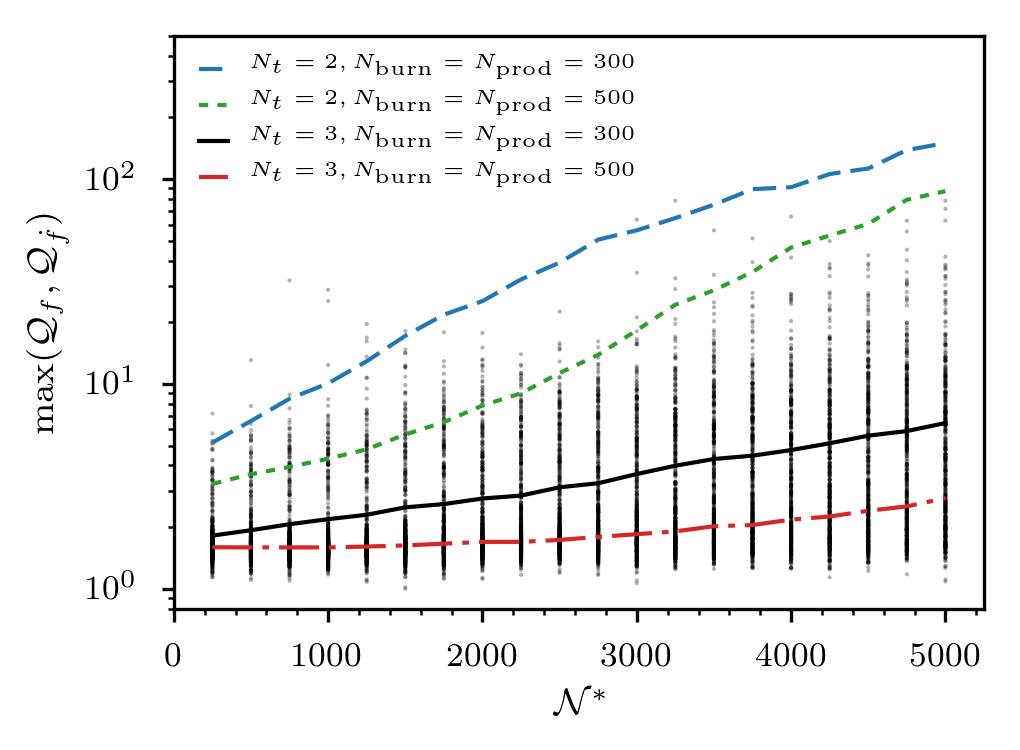}
\caption{Study of how $\mathcal{Q}$ (defined in Eq.~\eqref{eqn_Q}) depends on
the size of the prior parameter space \meta{(quantified in terms of $\N^*$)} and the search setups.
Four \ac{MC} studies are shown as line plots of the mean value. In each
of the four \ac{MC} studies, we vary the number of steps and the number of temperatures (the same
maximum temperature, see Section~\ref{sec_parallel_tempering}, is used throughout).
To illustrate the variance in a single \ac{MC} study, markers show the individual \ac{MC}
results for $\Ntemps=3$ and $\Nburn=\Nprod=300$, the default setup used in the rest of this work.}
\label{fig_convergence}
\end{figure}

Taking four different search setups (including the default setup used throughout this
work, $\Ntemps=3$ and 300 burn-in and production steps), we repeat the process
\CHECK{500} times in a \ac{MC} study. For each setup, we vary $\N^*$ and in
Fig.~\ref{fig_convergence} we plot the mean maximum of $\mathcal{Q}_f$ and
$\mathcal{Q}_{\dot{f}}$. For the default setup, we also plot the individual
results from the \ac{MC} study to illustrate the variability about the mean.

This figure illustrates that using a greater number of temperatures and steps
can improve the convergence efficiency. For the default setup, we
conservatively define $\N^*_\mathrm{max}\equiv\CHECK{1000}$ in order to ensure
that the $\mathcal{Q}\sim 1$ such that the sampler will converge

\section{Hierarchical multistage follow-up}
\label{sec_follow_up}

\begin{figure}[htb]
\includegraphics[]{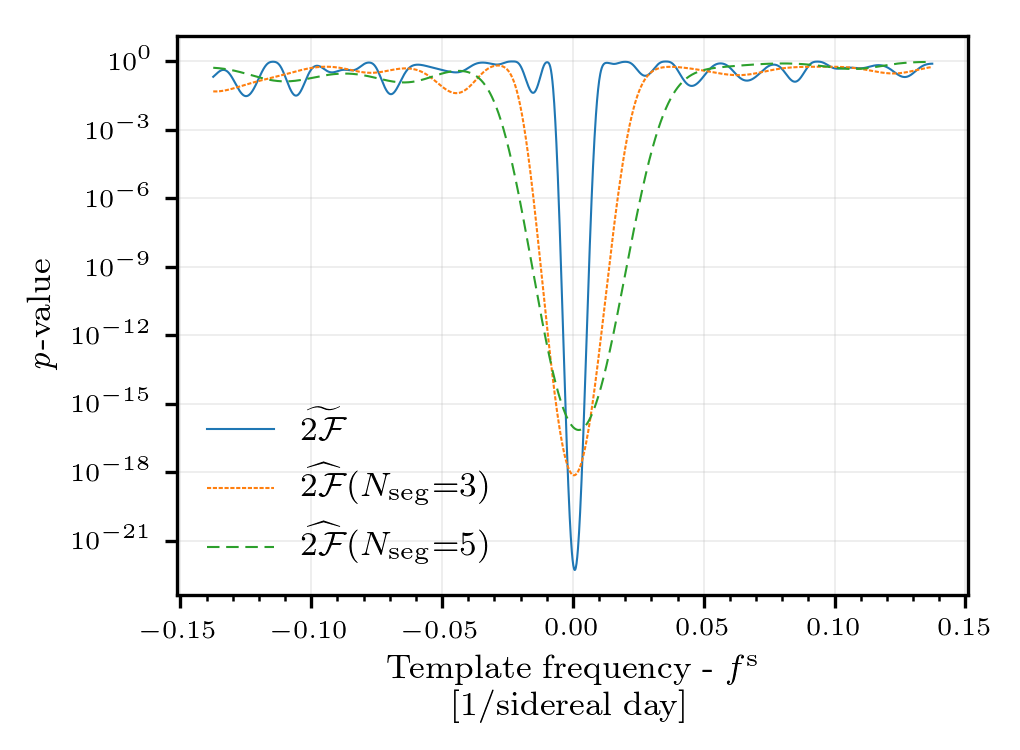}
\caption{Comparison of the fully coherent $\widetilde{2\mathcal{F}}$ and
semi-coherent $\widehat{2\mathcal{F}}$ as a function of the frequency for the
same simulated signal in noise.}
\label{fig_comparison}
\end{figure}

Semi-coherent detection statistics trade candidate significance for
computational cost to detect a candidate. To illustrate how, in
Fig.~\ref{fig_comparison}, we compare the significance, expressed as a
\emph{p}-value\footnote{Calculated by
computing the $\twoFtilde$ and $\twoFhat$ values at each frequency, and then
using that the background distribution for a fully coherent search in Gaussian
noise is $\chi^2$-distributed with 4 degrees of freedom while for a
semi-coherent search it is $\chi^2$-distributed with $4\Nseg$
degrees of freedom.}, as a function of frequency for a fully coherent and two
semi-coherent searches of a signal in Gaussian data. This demonstrates that
while the maximum significance of the fully coherent search is the largest (in that
its \emph{p}-value is the smallest of the three), the semi-coherent searches have
wider peaks. For a template bank search, using a semi-coherent detection
statistic means that fewer templates are required to cover the search space.

A semi-coherent search (e.g., an all-sky or directed search) may result in a list
of promising candidates; these are then processed in a \emph{hierarchical
follow-up} in which the number of segments is decreased from $\Nseg^{(0)}$, the
number used in the initial search, to $\Nseg^{(j_\mathrm{max})}{=}1$ (i.e.,
fully coherent). The ultimate aim of such a process is to either localize the
candidate using a fully coherent detection statistic (i.e., calculate the maximum
available significance) or refute the candidates as not being a standard
\ac{CW} if the detection statistic does not increase as expected.

The hierarchical follow-up can be done in two stages (cf.\ \citet{brady2000})
with an initial semi-coherent  stage followed directly by a fully coherent
search. However, it was shown in a numerical study by \citet{cutler2005} that a
multi-stage hierarchical follow-up search using a decreasing \emph{ladder} of
segment numbers $\{\Nseg^{(j)}\}$ can significantly improve the efficiency:
ultimately they concluded that for template bank searches three semi-coherent
stages provide the best trade-off between sensitivity and computational cost.
For further examples of follow-up searches, see \citet{shaltev2013} which used
a ``Mesh Adaptive Direct Search'' algorithm for optimization or \citet{papa2016,
abbott2017allsky, abbott2017allskyeinstein} for multi-stage gridded approaches.

In this work, we define a method to perform MCMC follow-up
searches; we will make use of the idea that in a semi-coherent MCMC search,
$\Nseg$ can be thought of as a free parameter which adjusts the width of signal
peaks (since the $\F$-statistic is proportional to our log-likelihood, cf.\
Eq.~\eqref{eqn_lambda_posterior_sc}).

The general principle is that, given a
ladder of segment numbers, we run the MCMC sampler for each stage of the
ladder, allowing enough time for the walkers to converge before moving to the next stage of the
ladder. At each new stage, the previously converged walkers will now be
overdispersed with respect to the new posterior distribution and will begin to
converge, provided the change in the number of segments is sufficiently small
so that they do not lose the signal. Repeating this process over the ladder
of coherence times, at each stage the signal is better localized until the final
fully coherent step\footnote{In some sense, this method bears a resemblance to
simulated annealing \citep{mackay2003information}, in which the likelihood is
raised to a power (the inverse temperature) and subsequently ``cooled''. For a
discussion and examples of using simulated annealing in the context of CW
searches see \citet{veitch2007}}. In the next Section, we define how to
optimally choose the ladder of segment numbers.

It is worth noting that we do not test for convergence after each stage. Should
the chains fail to converge, e.g., as in Fig.~\ref{fig_MCMC_non_convergence},
then it is likely subsequent stages will remain unconverged. A potential future
improvement is to implemented convergence checking, allowing stages to be
curtailed if convergence is reached before some maximum allowed number of
steps.

\subsection{Optimal ladder of coherence times}

Consider a candidate from a semi-coherent search at the $j$th stage with
$\Nseg^{(j)}$ segments and parameter uncertainty $\Delta\blambda^{(j)}$
(defined momentarily). The size of the signal within
that uncertainty can be quantified by writing Eq.~\eqref{eqn_Nstar} with an
explicit dependence on the number of segments and the parameter uncertainty,
\begin{align}
\N^*\left(\Nseg^{(j)}, \Delta\blambda^{(j)}\right) =
\sqrt{\mathrm{det}g\left(\Nseg^{(j)}\right)}\;\mathrm{Vol}(\Delta\blambda^{(j)})\,.
\end{align}
For an \ac{MCMC} search to be effective we need $\N^*(\Nseg^{(j)},
\Delta\blambda^{(j)}) \lesssim \N^*_\mathrm{max}$ (see
Section~\ref{sec_maximim_size_of_prior}).

For the $j=0$ stage, the parameter uncertainty, $\Delta\blambda^{(0)}$, depends on the
details of the initial
search. However, if the initial search was too coarse for the MCMC method to
be effective (i.e., $\Delta\blambda^{(0)}$ is too large), a template bank
refinement can first be performed
to reduce the uncertainty and ensure that $\N^*(\Nseg^{(0)}, \Delta\blambda^{(0)}) \lesssim
\N^*_\mathrm{max}$. If we then run the \ac{MCMC} search using
$\Nseg=\Nseg^{(0)}$ and a uniform prior based on $\Delta\blambda^{(0)}$, after
convergence the walkers will be approximately bound in a posterior parameter
space $\Delta\blambda^{(1)} \le \Delta\blambda^{(0)}$.  In a hierarchical
search, one then proceeds to increase the coherence time to $\Nseg^{(1)} <
\Nseg^{(0)}$ and again localize the signal. For our \ac{MCMC} search, we
similarly continue the \ac{MCMC} simulation, but with fewer segments,
better localizing the signal.

There are two consideration to be made in defining the ladder of coherence times.
Firstly, the change in $\Nseg$ between any two stages must be sufficiently
small such that the walkers converge to the new posterior, formally this means
that
\begin{align}
\N^*(\Nseg^{(j)}, \Delta\blambda^{(j)})
\le \N^*_\mathrm{max}\,.
\label{eqn_limit}
\end{align}
The second consideration is that we want to minimize the computational cost
of the follow-up. This is achieved by choosing the maximal allowable
decrease in $\Nseg$ while still respecting Inequality~\eqref{eqn_limit}.

The size of $\Delta\blambda^{(j+1)}$ is not known prior to
running the $j$th stage of the search. However, since
$\sqrt{g(\Nseg^{(j)})}$ is the coordinate size a
signal will occupy at a metric mismatch of one, (i.e., where the $\F$-statistic
or log-likelihood is small compared to the peak value), it is conservative to
assume that $\N^*(\Nseg^{(j-1)}, \Delta
\blambda^{(j)}) \approx 1$. Dividing this through
Inequality~\eqref{eqn_limit} and taking the equality to minimize the computational cost,
we see that one needs to find the largest $\Tcoh^{(j)}$ which satisfies
\begin{align}
\N^*_\mathrm{max} \approx \frac{\sqrt{g(\Tcoh^{(j)})}}{\sqrt{g(\Tcoh^{(j-1)})}}\,.
\end{align}
This can be done by increasing $\Tcoh^{(j)}$ in steps until this
criteria is satisfied; for efficiency, we instead define it as
a minimization problem and use standard numerical solvers. Given an
initial prior volume and coherence time, the optimal ladder of coherence times
can then be precomputed saving computation time during the run (useful as the
same setup can be applied to multiple candidates).

\subsection{Example}
\label{sec_followup_example}

We now provide an illustrative example of the follow-up method. We consider a
directed search in \CHECK{100} days of data from a single detector, with
$\sqrt{\Sn}=\CHECK{\DirectedFollowUpSqrtSn}$~Hz$^{-1/2}$ (at the fiducial
frequency of the signal). The simulated signal has an amplitude
$h_0=\DirectedFollowUphzero$ such that the signal has a sensitivity depth of
$\Depth=\DirectedFollowUpDepth$~Hz$^{-1/2}$ in the noise.  We choose arbitrary
phase-evolution parameters of $f\Signal=\CHECK{30}$~Hz,
$\dot{f}\Signal=\CHECK{10^{-10}}$~Hz/s, and the sky location of the Crab
pulsar. The initial prior uncertainty, $\Delta\blambda^{(0)}$, is $\Delta
f=\CHECK{\SCDirectedFUDeltaFzero}$~Hz and $\Delta
\dot{f}=\CHECK{\SCDirectedFUDeltaFone}$~Hz/s.
The optimal ladder of coherence times is precomputed using
$\N^*_\mathrm{max}=\CHECK{\DirectedFollowUpNstarMax}$ (i.e., the size of the
fully coherent parameter space) and the resulting setup
is given in Table~\ref{tab_directed_MC_follow_up}.

\begin{figure}[htb]
\centering
\includegraphics[width=0.5\textwidth]{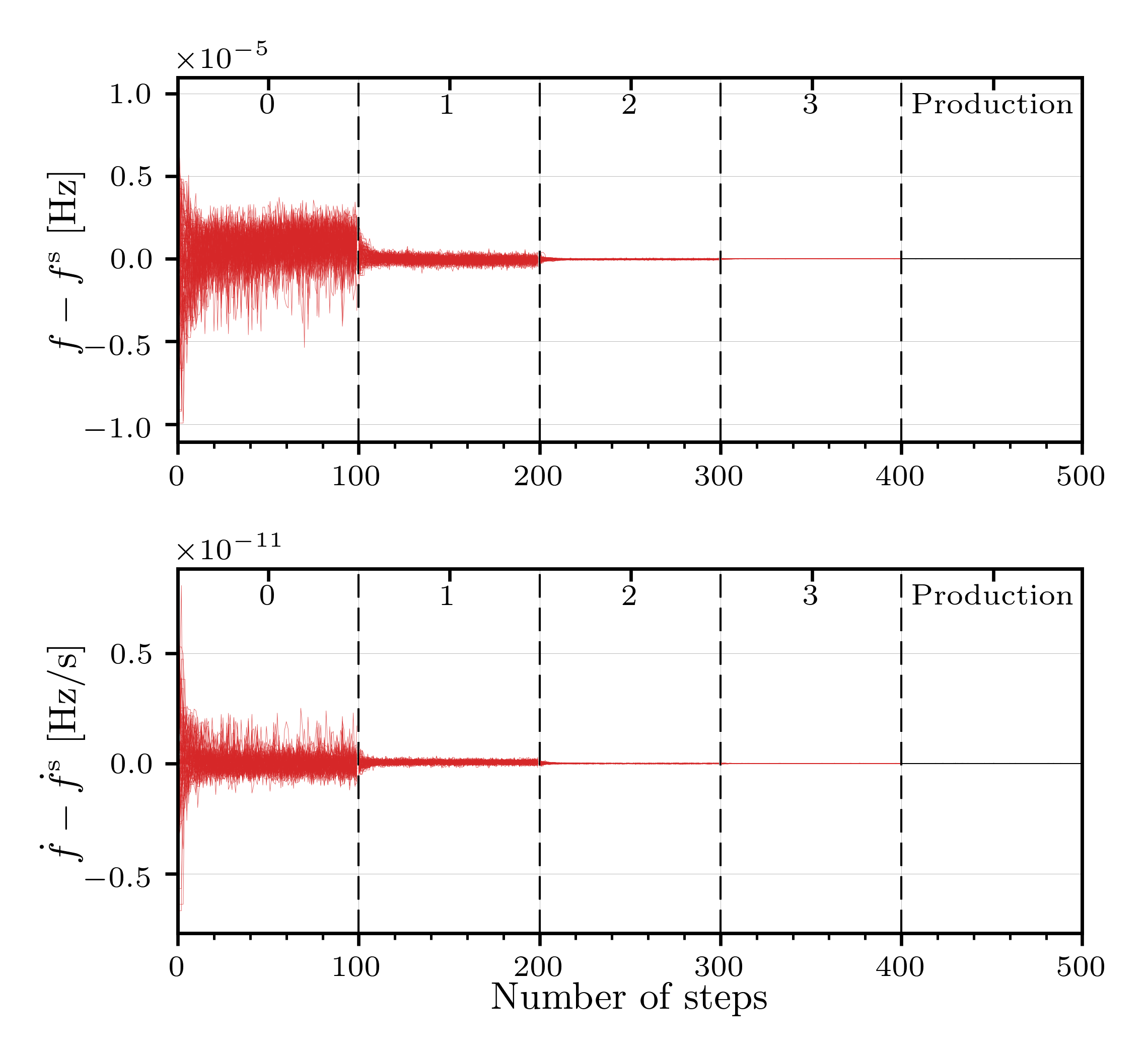}
\caption{We show the progress of the \CHECK{100} parallel
walkers (see Fig.~\ref{fig_MCMC_simple_example} for a description) during the
\ac{MCMC} simulation for the directed search parameters. Each vertical dashed
line indicates the start of a new stage of the search (numbered on the top
axis), the details of all stages are listed in Table~\ref{tab_directed_MC_follow_up}.}
\label{fig_follow_up}
\end{figure}

Fig.~\ref{fig_follow_up} shows the progress of the \ac{MCMC} sampler during
the follow-up. For illustrative purposes, in this example we use
$\CHECK{100}$ steps per-stage, rather that the default $300$ as suggested in
Section \ref{sec_example}. As expected from
Table~\ref{tab_directed_MC_follow_up}, during stage 0 the signal peak is broad
with respect to the size of the prior volume, therefore the \ac{MCMC}
simulation quickly converges to it. At each subsequent stage, when the number
of segments is reduced, the peak narrows and the samplers similarly converge to
this new solution.

\begin{table}[htb]
\caption{The search setup used in Fig.~\ref{fig_follow_up}, generated with
$\N^*_\mathrm{max}=\CHECK{\DirectedFollowUpNstarMax}$. The final column
provides the approximate number of unit-mismatch templates
at each stage calculated over the prior volume.}
\label{tab_directed_MC_follow_up}
\begin{tabular}{c|ccc}
Stage & $N_\mathrm{seg}$ &$T_\mathrm{coh}^{\rm days}$ &$\mathcal{N}^*(\Nseg^{(\ell)}, \Delta\boldsymbol{\lambda}^{(0)})$ \\ \hline
0 & 100 & 1.0 & 2.2 \\
1 & 35 & 2.9 & 18.0 \\
2 & 5 & 20.0 & $8.8{\times}10^{2}$ \\
3 & 1 & 100.0 & $1.0{\times}10^{4}$ \\
\end{tabular}

\end{table}

\subsection{Monte Carlo studies}
\label{sec_MCS}

In order to quantify how well the \ac{MCMC} follow-up method works, we test its
ability to successfully identify simulated signals in Gaussian noise. This will
be done in a \acf{MC} study, with independent random realizations of the
Gaussian noise. Such a method is analogous to the studies performed in
\citet{shaltev2013}, except that we present results as a function of the
injected sensitivity depth, rather than the squared-\ac{SNR}.

In Section~\ref{sec_directed_follow_up} and \ref{sec_all_sky_follow_up}, we
perform \ac{MC} studies for the follow-up of directed and all-sky searches. For both
studies, we simulate
an isotropic distribution of source orientations by drawing the amplitude
parameters for each signal uniformly from $\phi \in [0, 2\pi]$, $\psi \in
[-\pi/4, \pi/4]$, and $\cos\iota \in [-1, 1]$.  However, we do not draw $h_0$
randomly, but run the \ac{MC} study at a range of selected $h_0$ values to show
how the \ac{MCMC} follow-up performs as a function of sensitivity depth (we fix
$\sqrt{S_n}=\CHECK{1\times10^{-23}}$). The selection of phase-evolution
parameters is discussed separately for each study.

The success of an \ac{MCMC} follow-up search is evaluated by the maximum
$\twoFtilde$ value found in the final, fully coherent stage of the search. We
begin each \ac{MC} study by first estimating a background distribution for the
maximum $\twoFtilde$ value found in noise and using this to define a threshold
value $\twoFtilde_{\rm th}$; for the simulated signals in noise we then ascribe
results with a maximum $\twoFtilde$ above this threshold as ``recovered''
whilst those falling short are not recovered.

The hierarchical multi-stage \ac{MCMC} search may fail to detect a candidate
for two distinct reasons: (i) the simulated signal is not sufficiently loud
with respect to the background noise to be detected and (ii) the simulated signal
is sufficiently loud, but the \ac{MCMC} fails to recover it. The first reason
can be understood as saying that, even for a template bank search which covers
the search space with infinite resolution, for sufficiently weak signals, the
proportion of recovered candidates tends to zero. More precisely this behavior
is the expected \emph{theoretical optimal performance} detection probability
for an infinitely dense fully coherent search of data containing
isotropically-distributed signals as calculated by \citet{wette2012}. In this
study, we want to understand how the \ac{MCMC} follow-up performs with respect
to reason (ii).  Therefore, we will plot the recovered fraction (in the
\ac{MC} study) as a function of the sensitivity depth, and compare this against the
theoretical optimal performance (i.e., using Eq.~(3.8) of \citet{wette2012} to
relate the averaged-\ac{SNR} to the sensitivity depth).  Deviations of the
\ac{MC} study with respect to the theoretical optimal performance indicate any
weakness in the \ac{MCMC} follow-up method itself.

\subsubsection{Follow-up of candidates from a directed search}
\label{sec_directed_follow_up}

For the directed \ac{MC} study, we simulate signals following the example
described in Sec.~\ref{sec_followup_example}, except that the simulated
frequency and spin-down are selected randomly from the inner half of the
initial prior uncertainty. As in the example, the prior is defined to be
uniform over the full uncertainty box and the duration is 100 days. Therefore,
the optimal setup is also the same and given
in Table~\ref{tab_directed_MC_follow_up}. Each stage is run for 300 steps
before moving to the next stage of the ladder, after the final stage an
additional 300 production steps are taken.

Characterizing the search in Gaussian noise without a signal first, we simulate
$\CHECK{1\times10^{3}}$ realizations and perform the follow-up search on
these.  The largest observed value was found to be
$\CHECK{\DirectedMCNoiseOnlyMaximum}$. From this, we can set an arbitrary threshold for
the detection statistic of $\twoFtilde_{\rm th} = \CHECK{60}$.

\begin{figure}[htb]
\centering
\includegraphics[width=0.45\textwidth]{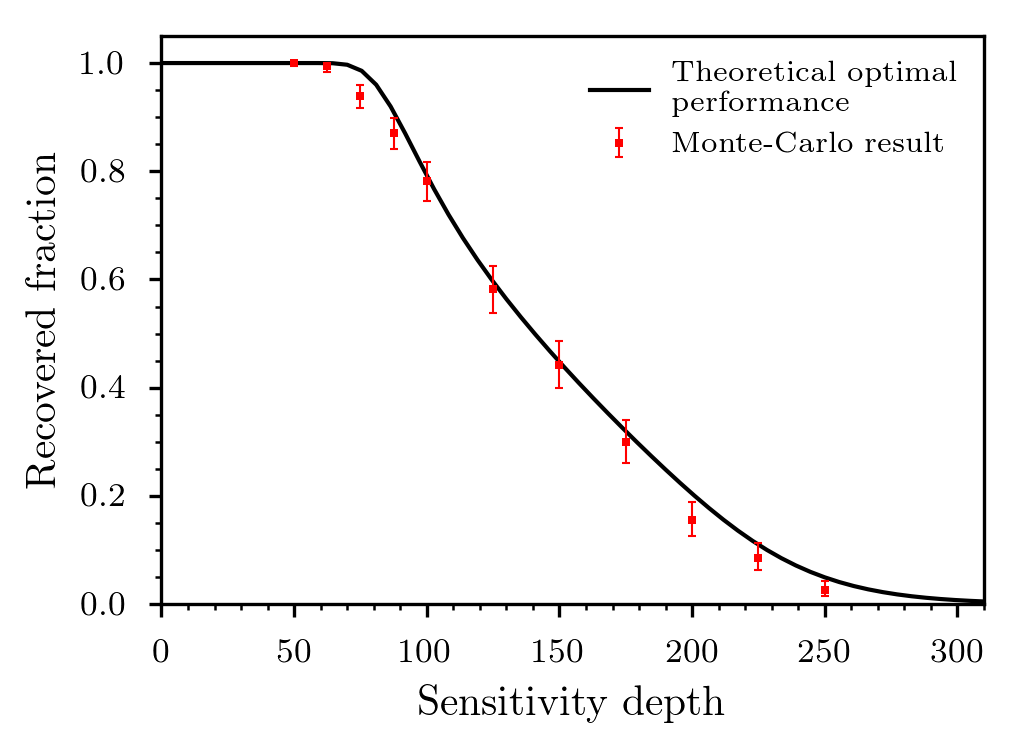}
\caption{Recovery fraction for the directed follow-up. The \ac{MC} results
come from random draws searches using the setup described in
Table~\ref{tab_directed_MC_follow_up}.}
\label{fig_directed_MC_follow_up}
\end{figure}

Running \CHECK{500} \ac{MC} simulations of Gaussian noise with
a simulated signal, in Fig.~\ref{fig_directed_MC_follow_up} we plot the
fraction of the \ac{MC} simulations which where recovered (i.e.,
$\twoFtilde^{\rm max} > \twoFtilde_{\rm th}=\CHECK{60}$) and compare this
against the theoretical optimal performance as discussed earlier, given the
threshold. The figure demonstrates that the recovery power of the \ac{MCMC}
follow-up shows negligible losses compared to the theoretical optimal
performance.

\subsubsection{Follow-up of candidates from an all-sky search}
\label{sec_all_sky_follow_up}

We now test the follow-up method when applied to candidates from an all-sky
search which, by definition, have uncertain sky-position parameters $\alpha$
and $\delta$. Searching over these parameters, in addition to the frequency
and spin-down not only increases the parameter space volume that needs to be
searched, but also adds difficulty due to correlations between the sky-position
and spin-down.

To replicate the condition of candidates from an all-sky search, we draw the
candidate positions isotropically from the unit sphere (i.e., $\alpha$ uniform
on $[0, 2\pi]$ and $\delta = \sin^{-1}(u)$ where $u$ is uniform
on $[-1, 1]$). We then place an uncertainty box containing the candidates with a
width $\Delta\alpha=\Delta\delta=\CHECK{0.05}$; this box is
chosen in such a way that the location of the candidate has a
uniform probability distribution within the box. This neglects the non-uniform
variation in $\delta$ over the sky-patch, but, given the small size of the
sky-patch, should not cause any significant bias. The frequency, spin-down, and
amplitude parameters are chosen in the same way as for the directed search
(Section~\ref{sec_directed_follow_up}). The optimal setup is precomputed for
this prior and given in Table~\ref{tab_allsky_MC_follow_up}. Each stage is run
for 300 steps before moving to the next stage of the ladder, after the final
stage an additional 300 production steps are taken.

\begin{table}[htb]
\caption{Run-setup for the all-sky follow-up \ac{MC} study, generated with
$\N^*_\mathrm{max}=1000$. The approximate number of unit-mismatch templates
over the initial search space was computed at the equator
(i.e., $\delta=0$) which produces the largest volumes.}
\label{tab_allsky_MC_follow_up}
\begin{tabular}{c|ccc}
Stage & $N_\mathrm{seg}$ &$T_\mathrm{coh}^{\rm days}$ &$\mathcal{N}^*(\Nseg^{(\ell)}, \Delta\boldsymbol{\lambda}^{(0)})$ \\ \hline
0 & 100 & 1.0 & 8.4 \\
1 & 35 & 2.9 & $5.6{\times}10^{2}$ \\
2 & 6 & 16.7 & $5.2{\times}10^{5}$ \\
3 & 1 & 100.0 & $2.5{\times}10^{7}$ \\
\end{tabular}

\end{table}

Again, we first characterize the behavior of the all-sky follow-up by applying
it to $\CHECK{1\times10^3}$ realizations of Gaussian noise.
%The resulting histogram is provided in Fig.~\ref{fig_hist_AllSkyMCNoiseOnly} and
The largest $\twoFtilde$ value was found to be $\CHECK{\AllSkyMCNoiseOnlyMaximum}$. This is larger
than the value found for the directed search, although both use the same number
of Gaussian noise trials, and therefore must result from an increased number
of independent templates. As a result we correspondingly increase our 
arbitrary detection threshold for the all-sky search to $\twoFtilde_{\rm tr} =
\CHECK{70}$.

\begin{figure}[htb]
\centering
\includegraphics[width=0.45\textwidth]{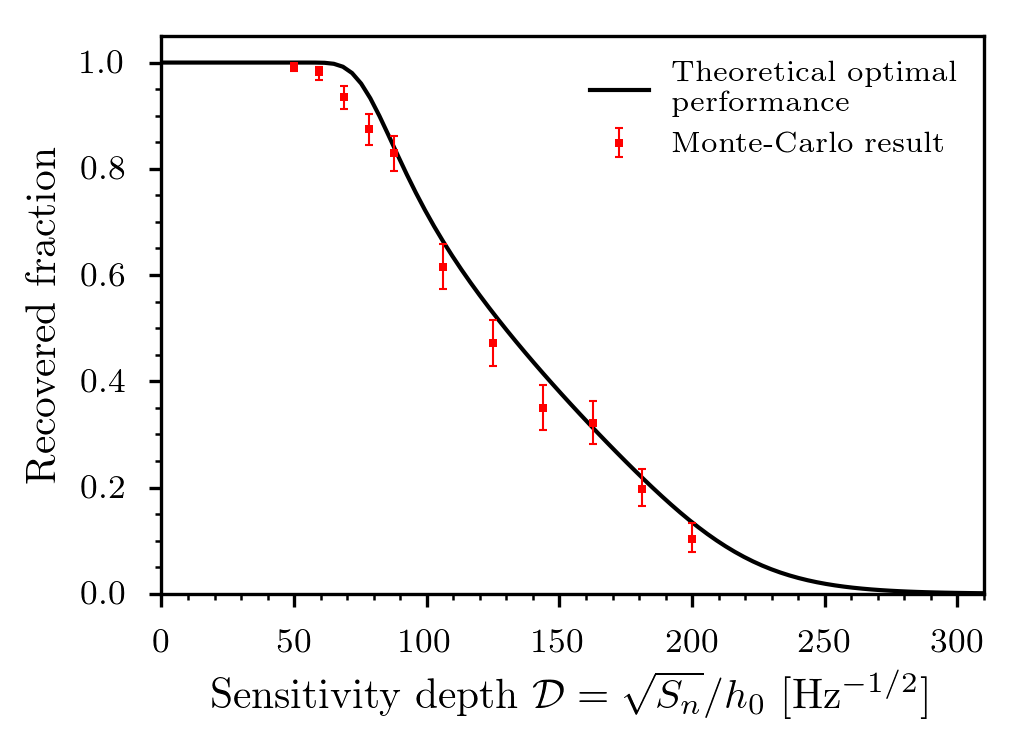}
\caption{Recovery fraction for the all-sky follow-up. The \ac{MC} results
come from random draws searches using the setup described in
Table~\ref{tab_allsky_MC_follow_up}.}
\label{fig_allsky_MC_follow_up}
\end{figure}

Running \CHECK{500} \ac{MC} simulations of Gaussian noise with randomly drawn
signals, the resulting recovery fraction as a function of the injected
sensitivity depth is given in Fig.~\ref{fig_allsky_MC_follow_up}.  We find
that the all-sky Monte Carlo has a detection efficiency close to the
theoretical optimal performance.

\section{Computation time estimates}
\label{sec_computing_cost}

In this section, we will provide an approximate timing model for the MCMC
search method implemented in \textsc{pyfstat}.

The computation time of an \ac{MCMC} search can be split into two
contributions, the evaluation of the $\F$-statistic and all other operations,
which we refer to as the \textsc{pyfstat}-overhead
(i.e., computing the prior probabilities, generating the proposal jumps,
evaluating the jumps and other minor contributions).

Timing the evaluation of the ``Demod'' $\F$-statistic (used in this analysis)
is discussed in \citet{prix2017}. The time to compute a
single-$\blambda$ template for a single detector can be written as
$\tauFeff=\Nsft\tauLD$, where $\Nsft$ is the number of short Fourier transforms
(SFTs) (the core data product used in \ac{CW} searches), which will depend on
the data duration and SFT length (typically 1800~s). We can further
decompose $\tauLD$ as
\begin{align}
\tauLD = \tauCore + b\tauBuffer\,,
\end{align}
where $\tauCore$ is the core time to compute a single
$\F$-statistic evaluation (excluding the time to compute any buffered
quantities), $\tauBuffer$ is the time to compute the buffered quantities, and
$b \in [0, 1]$ the ``buffer miss fraction'' quantifies how often the buffer
needs to be recomputed between evaluations \citet{prix2017}. For our purposes, it is sufficient
to note that for all-sky searches $b=1$ (if the sky position changes between
steps), but for directed searches~$b\approx1$.

The \textsc{pyfstat}-overhead will depend on whether the search is fully or
semi-coherent. We now discuss each of these in turn.

\subsection{Fully coherent timing model}
For any fully coherent search using the \textsc{pyfstat} code, the time
per-call is given by
\begin{align}
\tau_\mathrm{FC} = (\tauS + \tauLD \Nsft)\,,
\end{align}
such that the total time for a fully coherent MCMC search with $\Nwalkers$ walkers
and $\Ntemps$ temperatures is
\begin{align}
\mathcal{T}_\mathrm{FC}^\mathrm{MCMC} = \tau_\mathrm{FC} (\Nburn+\Nprod)\Nwalkers\Ntemps\,.
\label{eqn_fc_timing}
\end{align}
Note that this timing model is independent
of the state of the \ac{MCMC} sampler, i.e., whether or not it has converged to a
single peak.

To estimate these timing coefficients, we run \ac{MC} studies of \ac{MCMC}
searches for various data spans ranging from \CHECK{25}~days to \CHECK{100}~days. Profiling the
run times, we estimate $\tauLD$ from the calls to \textsc{XLALComputeFstat()},
then $\tauS$ as the remaining run time. The resulting estimates are given in
Table~\ref{tab_timing}.

\begin{table*}
\centering
\caption{Estimates of the fundamental timing coefficients made on a Lenovo
T460p with an Intel(R) Core(TM) i5-6300HQ CPU @ 2.30GHz.}
\label{tab_timing}
\begin{tabular}{|l|l|c|c|c|c|} \cline{3-6}
\multicolumn{2}{l|}{} & $\tauLD$ [s] & $\tauT$ [s] & $\tauS$ & $\tauSCeff$ [s] \\ \hline
\multirow{2}{*}{Directed} & fully coherent
    & $7.4\times10^{-8}\pm27\%$
    & ---
    & $6.4\times10^{-5} \pm 14\%$
    & --- \\ \cline{2-6}
& semi-coherent
    & $1.3\times10^{-7}\pm35\%$
    & $1.5\times10^{-8}\pm6\%$
    & $9.1\times10^{-5}\pm23\%$
    & $5.5\times10^{-6}\pm14\%$ \\
\hline
\multirow{2}{*}{All-sky} & fully coherent
    & $5.0\times10^{-7}\pm14\%$
    & ---
    & $1.0\times10^{-4} \pm 19\%$
    & --- \\ \cline{2-6}
& semi-coherent
    & $5.2\times10^{-7}\pm18\%$
    & $1.5\times10^{-8}\pm6\%$
    & $1.2\times10^{-4}\pm25\%$
    & $5.8\times10^{-6}\pm19\%$\\\hline
\end{tabular}
\end{table*}

The estimated $\tauLD$ differs between the all-sky and directed searches since
$b=1$ in the former, but $0$ for the latter. The estimated values of $\tauLD$
are in agreement with values found using gridded searches on the same machine,
and the values found in \citet{prix2017}. From this table, we see that
the evaluation of the $\F$-statistic will dominate the
timing when $\Nsft \gtrsim 900$  $(200)$, which for 1800\,s SFTs corresponds to $\sim
18$ $(4)$\,days for the directed (all-sky) search.

The dominant factor required to determine the cost of a run is the number of
steps (typically one uses a few hundred walkers and two or three temperatures).
The number of steps depends on the \ac{ACT}. For the directed
\ac{MC} study we found the largest \ac{ACT} to be
$\sim15$. To safely account for burn-in we therefore suggest the total number of
steps be $\sim 20\times15=300$.  Meanwhile, for the all-sky \ac{MC} study we found the
largest \ac{ACT} to be $\sim90$, so an all-sky search
requires a factor of $6$ more steps. These numbers are approximate and will
depend on the other search setup parameters (i.e., using only a single
temperature will require a larger numbers of steps); a \ac{MC}
simulation could be run prior to a follow-up to tune $\Nwalkers, \Ntemps$, and
the number of required burn-in and production steps.

\subsection{Semi-coherent timing model}

The current implementation of \textsc{pyfstat} implements a semi-coherent search
by using the \textsc{ComputeTransientFstatMap()} functions of
\textsc{LALSUITE}. This is done to avoid repeated computation of intermediate
data products, but itself incurs additional computational cost proportional
to the number of segments. We can model this by introducing a new
timing coefficient $\tauT$ such that the semi-coherent timing per-call is
\begin{align}
\begin{split}
\tau_\mathrm{SC} & = \left[
\tauS + \tauSCeff\Nseg
+ \left(\tauLD + \tauT\Nseg\right)
\Nsft\right]\,,
\end{split}
\end{align}
with the total time being
\begin{align}
\tau^\mathrm{MCMC}_\mathrm{SC} = \tau_\mathrm{SC} (\Nburn+\Nprod)\Nwalkers\Ntemps\,.
\label{eqn_sc_timing}
\end{align}
Again running a \ac{MC} study, we present
estimates of these timing constants in Table~\ref{tab_timing}.

\subsection{Follow-up timing model}

The computing cost for a given follow-up run setup can be estimated by summing
the cost for each stage using Eq.~\eqref{eqn_fc_timing} and
\eqref{eqn_sc_timing}. As an example, the directed search run setup given in
Table~\ref{tab_directed_MC_follow_up} has an estimated run time of
$\sim\CHECK{432}$~s while the all-sky run setup in
Table~\ref{tab_allsky_MC_follow_up} has an estimated run time of $\sim\CHECK{2200}$~s.
The limitation of this method is that it requires the candidate to be
sufficiently well localized at the coherence time of the original search.
Therefore, an additional refinement stage may be required prior to the MCMC
follow-up which will add an additional overhead.

\section{Conclusion}
\label{sec_conclusion}

We investigate the use of an \ac{MCMC} method in continuous-gravitational-wave
searches. Compared to template-bank searches, these enable efficient
exploration of the parameter space and posterior parameter inference.  However,
these advantages may only be reaped when the signal occupies a
reasonably large fraction of the prior volume. We quantify this through $\N^*$, the
approximate number of unit-mismatch templates \addtext{required to cover the prior volume}.
\addtext{In particular,
the prior volume must be sufficiently small in the sense that $\N^*\lesssim \N^*_\mathrm{max}$.
For this reason, \ac{MCMC} methods are generally not suitable as an initial search method
in wide parameter spaces where this condition is not met.}

\addtext{We have investigated $\N^*_\mathrm{max}$ for the \texttt{ptemcee}
sampler with a particular setup (see Sec.~\ref{sec_parallel_tempering}) and
fixed number of steps, and found $\N^*_{\mathrm{max}}\sim1000$ to be a
reasonably safe choice. In general, this value will depend on the sampler,
setup, and number of steps. However, we expect our general conclusion to hold for
different samplers (including nested sampling methods), but $\N^*_\mathrm{max}$
could differ. In the future we hope to extend this work to use a variety of
samplers, in which case $\N^*_\mathrm{max}$ will provide a method to compare
their efficiency.}

We furthermore propose an MCMC-based hierarchical multi-stage follow-up for
signal candidates found in semi-coherent wide-parameter-space searches. The MCMC
methods are again found to be suitable in this context.  We define a method to determine the
optimal setup, balancing computational efficiency against robustness to losing
signals. Testing against the theoretical optimal performance (i.e., an
infinitely fine template bank), we find that the loss of signals inherently
due to the MCMC procedure is small.

\section{Acknowledgments}
We thank David Keitel, Simone Mastrogiovanni, Grant Meadors, and Karl Wette for
useful comments during the development of this work.

\appendix

\def\mnras{Mon. Notices Royal Astron. Soc.}
\def\cqg{Class. Quantum Grav.}
\bibliography{bibliography}

\end{document}